\documentclass[%
 reprint,
superscriptaddress,
amsmath,amssymb,
aps,
prb,
floatfix,
]{revtex4-2}

\usepackage[colorlinks]{hyperref}
\usepackage{amsmath}
\usepackage{adjustbox}
\usepackage{graphicx}
\usepackage{dcolumn}
\usepackage{bm}

\usepackage[svgnames]{xcolor}

\usepackage{physics}
\usepackage{multirow}

\newcommand\sro{Sr$_2$RuO$_4$}
\newcommand\sio{Sr$_2$IrO$_4$}
\newcommand\bio{Ba$_2$IrO$_4$}
\newcommand\bro{Ba$_2$RhO$_4$}

\newcommand\jeff{$j_{\mathrm{eff}}$}
\newcommand\jstar{$j_{\mathrm{eff}}^*$}
\newcommand\vk{\mathbf{k}}
\newcommand\wf{\mathcal{G}_{0}}
\newcommand\gloc{G_{\mathrm{loc}}}
\newcommand\gimp{G_{\mathrm{imp}}}
\newcommand\ham{\hat{\mathrm{H}}}
\newcommand{\vR}[1]{\mathbf{R_{#1}}}

\newcommand{\vK}[1]{\mathbf{k_{#1}}}

\newcommand{\anh}{\hat{c}^{\phantom{\dagger}}}
\newcommand{\cre}{\hat{c}^{\dagger}}
\newcommand{\ope}[1]{\hat{#1}}

\begin{document}

\preprint{APS}

\title{
 Affordable Five-Orbital Dynamical Mean-Field Theory for Layered Iridates and Rhodates 
}

\author{Léo Gaspard}
 \email{leo.gaspard@outlook.fr}
\affiliation{%
Laboratoire de Chimie et Physique Quantiques, Université de Toulouse, CNRS UMR 5626, 118 Route de Narbonne, 31062 Toulouse Cedex 09, France
}%
\affiliation{
Institut de minéralogie, de physique des matériaux et de cosmochimie, Sorbonne Université, CNRS UMR 7590, 4 Place Jussieu, F-75005 Paris, France
}

\author{Cyril Martins}%
\affiliation{%
Laboratoire de Chimie et Physique Quantiques, Université de Toulouse, CNRS UMR 5626, 118 Route de Narbonne, 31062 Toulouse Cedex 09, France
}%

\date{\today}

\begin{abstract}
\begin{description}
\item[Abstract]
Full $d$-manifold DMFT with numerically exact solvers has remained computationally prohibitive for spin-orbit materials due their scaling and severe sign problem, forcing the community to rely on simplified one- and three-band models that omit the $e_g$ states despite their proximity with the $t_{2g}$ orbitals.
We present the first full five-orbital Dynamical Mean-Field Theory (DMFT) calculations including spin-orbit coupling for the layered iridates and rhodates \bio~and \bro, revealing that the correlation effects shift significantly the $e_g$ states through static mean-field corrections rather than dynamical fluctuations. 
Motivated by this insight, we introduce hybrid-DMFT (hDMFT), which treats these orbitals and their coupling to the low-energy manifold at the mean-field level while maintaining near quantitative accuracy at a drastically reduced computational cost. 
These calculation establish hDMFT as a practical and accurate method for full $d$-manifold studies of layered iridates and rhodates, enabling systematic investigations of temperature, doping and pressure dependence that were previously computationally intractable.

\end{description}
\end{abstract}

\maketitle


\section{Introduction}
In recent years, transition-metal oxides with strong spin-orbit coupling (known as spin-orbit materials) have emerged as a fertile ground for the discovery of exotic electronic phases.
Their interplay between electron correlations, crystal-field effects and spin-orbit interaction gives rise to unconventional magnetism \cite{rauannu.rev.condens.matterphys.2016,jackeliphys.rev.lett.2009,winterj.phys.:condens.matter2017,thakurz.furanorg.allg.chem.2024,loveseyphys.rev.b2020}, spin-orbit Mott insulating behavior \cite{kimphys.rev.lett.2008,kimscience2009,mohitkarinorg.chem.2016,shinaokaphys.rev.lett.2015} and proximity to superconductivity \cite{kimscience2014,caonatcommun2016,schuckphys.rev.b2006}. 
Among these materials, layered $4d$ and $5d$ compounds such as iridates and rhodates have drawn particular attention \cite{martinsj.phys.:condens.matter2017}. 
These materials combine a quasi two-dimensional crystal structure \cite{crawfordphys.rev.b1994} with a bandwidth, Coulomb interaction strength, and spin-orbit coupling strength that all lie 
within the same order of magnitude \cite{mattheissphys.rev.b1976,pesinnaturephys2010},
making them paradigmatic systems for studying the intricate physics emerging from their interplay. 

Notably, the Ruddlesden-Popper series including strontium and barium iridates (\sio, \bio) and their rhodate analogues (Sr$_2$RhO$_4$, \bro) shares a structural proximity with the superconducting ruthenate Sr$_2$RuO$_4$ and the famous high-temperature superconducting copper oxides of the La$_2$CuO$_4$ family, suggesting that related emergent phenomena could occur in these compounds as well, even if no superconducting phase has yet been observed in these materials. 
\begin{figure*}[htp]
    \centering 
    \includegraphics[width=0.95\linewidth]{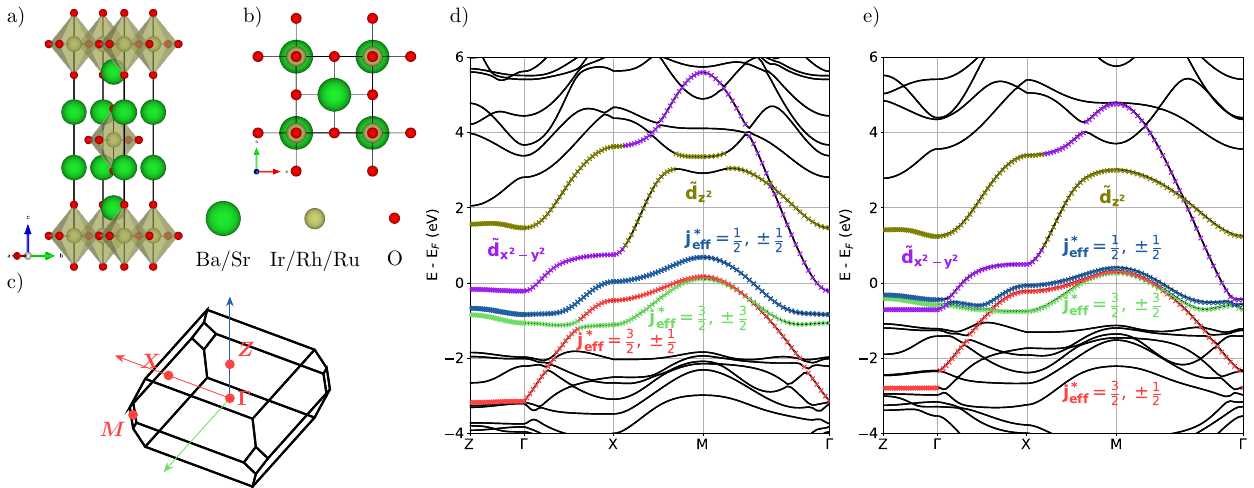}
    \caption{a) Conventional unit cell of \bio~and \bro. b) Cut of one IrO$_2$ or RhO$_2$ layer. c) First brillouin zone with the high-symmetry points used in this article. $\Gamma = (0, 0, 0)$, $X = (\pi, 0, 0)$, $M = (\pi, \pi, 0)$, $Z = (0, 0, \pi)$. d-e) Kohn-Sham band structure of \bio\ (d) and \bro\ (e) computed with spin-orbit coupling using the PBE exchange-correlation functional and a PW+ONCVPSP basis. For both materials, four bands cross the Fermi level : the three $t_{2g}$-like ones and the $d_{x^2-y^2}$ band.}
\label{fig:fig1}
\end{figure*}
Theoretical studies of
layered iridates and rhodates have commonly employed Density Functional Theory with a Hubbard correction (DFT+U) \cite{kimphys.rev.lett.2008,mosernewj.phys.2014,liu.phys.rev.lett.2008} or the more advanded Dynamical Mean-Field Theory (DMFT) which allows an explicit treatment of local dynamical fluctuations \cite{metznerphys.rev.lett.1989,vollhardtphysicab:condensedmatter1991,georgesrev.mod.phys.1996}, 
in the so-called DFT+DMFT approach\cite{martinsphys.rev.lett.2011,aritaphys.rev.lett.2012,zhangphys.rev.lett.2013,hampel.phys.rev.B.2015,moutenet.phys.rev.B.2018,martinsphys.rev.mater.2018,lenzj.phys.:condens.matter2019},

In these studies, spin-orbit coupling (SOC) can be treated either at the same level as the dynamical fluctuations or separately, and it was recently shown \cite{louatphys.rev.b2019} that treating SOC at the same level of theory as the electronic correlation was crucial for an accurate description of the materials. 
However, when treated at the same level of theory as the dynamical fluctuations, the state of the art DMFT solver, the numerically exact Continuous Time Quantum Monte-Carlo (CT-QMC) \cite{gullepl2008,wernerphys.rev.lett.2006} can only be applied to systems of less than five orbitals for a reasonable computational cost. 
This limitation has led the community to study layered iridates and rhodates using simple one- \cite{lenzj.phys.:condens.matter2019,martinsphys.rev.mater.2018} and three-band \cite{martinsphys.rev.lett.2011,perkinsphys.rev.b2014,zhangphys.rev.b2021,hampel.phys.rev.B.2015,zhangphys.rev.lett.2013} models, which excluded the $e_g$ bands from the description. 

Such minimal models were "historically" justified by
the large crystal field induced by the octahedral oxygen environment around the metal 
which i) splits the $d$ states into a partially filled $t_{2g}$ manifold and a (mostly) empty $e_g$ manifold.
and ii) 
allows to neglect the effect of spin-orbit coupling between the $t_{2g}$ and $e_g$ manifolds.
This last approximation called \textit{TP-equivalence approximation} \cite{sugano1970} leads to the so-called \jeff=$3/2$ and \jeff=$1/2$ states in the $t_{2g}$ subspace, while the $e_g$ one remains unaffected by the spin-orbit coupling. The so-called \jeff~picture was verified experimentally in \sio\ and \bio\ \cite{mosernewj.phys.2014,kimphys.rev.lett.2008,boseggiaj.phys.:condens.matter2013}.
More recent calculations on these compounds \cite{cassolphys.rev.b2024,zhangphys.rev.b2021} have taken into account further splitting within the $t_{2g}$ due to the tetragonal distortions of the octahedron along the $z$ axis, combined with the different chemical environment around the apical and basal oxygen atoms, but the \jeff~picture still remains the standard description for these compounds.

In this article, we claim that a full $d$-manifold model is the most "natural" description for layered iridates and rhodates, based on theoretical, numerical and experimental arguments. That is why, we developed a modification of the DMFT
loop which allows the full $d$-manifold DMFT-calculations for these materials at a drastically reduced computational cost, while maintaining near quantitative accuracy. 
We coin this method hybrid-DMFT (hDMFT for short). In a nutshell, hDMFT proposes a calculation in the $t_{2g}$ manifold embedded in a mean-field treatment of the $e_g$ subspace and of the $t_{2g}-e_g$ coupling. This approach is inspired by the DMFT+Hartree procedure \cite{hansmannnewj.phys.2014, raiphys.rev.x2024, dattanatcommun2023} in which the impurity problem is split into layers that are solved with different solvers, and we argue that such a mean-field treatment is justified by the TP-equivalence and the expected emptiness of the $e_g$ states.

Our study will focus on the two undistorted compounds of the iridate/rhodate family, namely \bio\ and \bro .  
Barium iridate, \bio, which is an in-plane antiferromagnetic insulator below 240K \cite{okabephys.rev.b2011} and was found to remain insulating up to at least 300K
\cite{nicholsappliedphysicsletters2014,mosernewj.phys.2014}, 
has already been studied theoretically 
by DFT+U \cite{uchidaphys.rev.b2014,mosernewj.phys.2014} and DFT+DMFT \cite{cassol2025,cassolphys.rev.b2024,aritaphys.rev.lett.2012}, 
suggesting a half-filled \jeff$= 1/2$ ground state. 
Barium rhodate \bro, which was recently synthesized using a high-pressure technique \cite{kurataphys.rev.materials2021}, 
behaves as a Fermi-liquid from 50K to at least 300K. 
No superconducting transition was observed through resistivity measurements as low as 160 mK but  
the importance of electronic correlations in this material were highlighted through specific-heat measurements, making a DMFT study of this compound relevant. The interest of studying these two compounds is double : i) a DFT+DMFT calculation really requires a five-orbital model for these materials and ii) their smaller unit cell allows these calculations (even without hDMFT) to remain computationally affordable, albeit with some effort.

The remainder of this article is organized as follows. First, we will introduce the local five-orbital model for layered iridates and rhodates and explained how it can be extracted from DFT calculations. We will also expose several arguments in favor of full-$d$ DMFT calculations for these materials. Then, we will introduce the hybrid-DMFT method that we developped to make such calculations possible.
Finally, we will present the five-band DMFT calculation for \bio~and \bro , and systematically compare the results from 
a full five-orbital DMFT calculation and the same five-orbital calculation using hDMFT, highlighting
the advantages of treating the $e_g$ states and the $e_g$-$t_{2g}$ coupling at the mean-field level. 

\section{Five orbital model for layered iridates and rhodates}

\subsection{Low-energy model}

In \autoref{fig:fig1}, the crystal structure and the Kohn-Sham band structure of \bio\ and \bro\ computed with SOC and using the PBE exchange-correlation functional and a PW+ONCVPSP basis are depicted.
For both \bio\ and \bro , the Kohn-Sham (PBE+SOC) 
description 
(see \autoref{fig:fig1} d) and e)) places the $d_{x^2-y^2}$ band below the Fermi level around the $\Gamma$ point. To perform a DFT+DMFT calculations on these compounds, the most "natural" low-energy model cannot therefore be restricted to the $t_{2g}$ manifold and should include the $e_g$ states too.  
We then chose to model \bio\ and \bro\ through an extended Hubbard-Kanamori Hamiltonian \cite{hubbardproc.r.soc.lond.ser.a1963,kanamoriprogressoftheoreticalphysics1963} of the following form : 

    \begin{equation}
        \ham = \ham_0 + \ham_{\mathrm{SOC}} + \ham_{\mathrm{int}}
    \end{equation}

    with the one-body term defined by :
    \begin{equation}
            \label{eq:model_kin}
            \ham_0 = \sum_\sigma \sum_{i\vR{}, j\vR{}'} t^{\phantom{\dagger}}_{i\vR{}, j\vR{}'} \cre_{\sigma i\vR{}^{\phantom{\prime}}}\anh_{\sigma j\vR{}'}
    \end{equation}
    The indices $\vR{},\vR{}'$ run over the metal sites and $\sigma$ denotes the electron spin. The labels $i, j$ refer to the five $d$ orbitals on each site. $\cre_{\sigma j\vR{}'}$ ($\anh_{\sigma j\vR{}'}$) are electron creation (annihilation) operators and $t^{\phantom{\dagger}}_{i\vR{}, j\vR{}'}$ are the tight-binding (TB) parameters. Local on-site energies are included in the kinetic term for $\vR{}=\vR{}'$.

    The Coulomb interaction $\ham_{\mathrm{int}}$ is also local :
    \begin{align}
            \label{eq:model_int}
            \ham_{\mathrm{int}} =  &\frac{1}{2} \sum_{\sigma} \sum_{ij} U_{ij} \hat{n}_{i\sigma}\hat{n}_{j\bar{\sigma}} + \frac{1}{2} \sum_{\sigma} \sum_{i \neq j} (U_{ij} - J_{ij}) \hat{n}_{i\sigma}\hat{n}_{j\sigma} \nonumber \\
                                   &- \frac{1}{2} \sum_{\sigma}\sum_{i\neq j} J_{ij} \left[ \cre_{i\sigma}\anh_{i\bar{\sigma}}\cre_{j\bar{\sigma}}\anh_{j\sigma} - \cre_{i\sigma}\cre_{i\bar{\sigma}}\anh_{j\sigma}\anh_{j\bar{\sigma}}\right]
    \end{align}
    The first two terms are the density-density terms, representing the Coulomb repulsion between electrons with antiparallel and same spin respectively.
    The last one includes the spin-flip and pair hopping terms. For the sake of readability, we omitted the sum over $\mathbf{R}$ in the expression of $\ham_{\mathrm{int}}$.

    Finally, in the Russel-Saunders coupling scheme, the spin-orbit term $\ham_{\mathrm{SOC}}$ is assumed to be purely local. 
    The SOC constant can be considered anisotropic \cite{zhangphys.rev.lett.2016} but we will in this article consider it to be purely isotropic and equal to $\mathbf{\lambda}$. Between the spin-orbitals $(i, \sigma)$ and $(j, \sigma')$, where $i,j$ represent the orbital indices and $\sigma$ the spin index, the spin-orbit coupling Hamiltonian is then :

\begin{equation}
    \ham_{\mathrm{SOC}} = \lambda \sum_{\sigma, \sigma'}\sum_{i,j, \vR{}} \mel**{i\sigma}{\mathbf{\ope{L}}\cdot\mathbf{\ope{S}}}{j\sigma'} \cre_{\sigma i \vR{}}\anh_{\sigma'j\vR{}}
    \label{eq:hamSO}
\end{equation}
where $\vR{}$ represents the site index.

The parameters of $\ham_0$ are computed using Maximally Localized Wannier Functions \cite{marzariphys.rev.b1997,souzaphys.rev.b2001} and the parameters of $\ham_{\mathrm{int}}$ are computed using constrained Random Phase Approximation (cRPA)\cite{aryasetiawanphys.rev.b2004}. We used the implementation of these methods available in the code RESPACK \cite{nakamuracomput.phys.commun.2021}. The value of the spin-orbit coupling strength is computed using a least-square fitting of the Wannier hamiltonian with spin-orbit interaction to the DFT+SOC band structure following the procedure outlined in \cite{gucomputationalmaterialsscience2023}. All the technical details for these calculations can be found in 
\autoref{app:computational}. 

The spin-orbit coupling strength ($\lambda$) and the average value of the Coulomb ($\bar{U}$) and exchange ($\bar{J}$) matrices are reported in \autoref{tab:param}, a more detailed description of the parameters of the models can be found in \autoref{app:models}. 

\begin{table}
    \centering 
    \begin{tabular}{|c|c|c|c|c|c|}
        \hline 
        & $\Delta$ (eV)& $\delta$ (eV)& $\lambda$ (eV)& \parbox[c][15pt][c]{1.1cm}{$\bar{U}$ (eV) }& $\bar{J}$ (eV)\\ 
        \hline
        \bio & 3.14 & 0.24 & 0.31 & 2.25 & 0.24  \\
        \bro & 2.60 & 0.22 & 0.09 & 1.73 & 0.24 \\ 
        \hline 
    \end{tabular}
    \caption{Crystal field splitting ($\Delta$), tetragonal splitting ($\delta$), spin-orbit coupling constant $\lambda$, average local Coulomb interaction ($\bar{U}$) and average exchange interaction ($\bar{J}$) 
    computed from first-principles for \bio~and \bro.}
    \label{tab:param}
\end{table}

\subsection{Local five-orbital hamiltonian and TP-equivalence approximation}

In layered iridates and rhodates, the local chemical environment of the Ir/Rh atoms (see \autoref{fig:fig1} a) and b)) induces 
a crystal field splitting $\Delta$ and a tetragonal splitting $\delta$ ($\delta'$) in the $t_{2g}$ ($e_g$) states. Taking into account the spin-orbit coupling of strength $\lambda$, the one-body local Hamiltonian matrix for the $d$ orbitals consists of two independent blocks that can be written 
in the bases $\{ d_{xz} \pm, d_{yz} \pm, d_{xy} \mp, d_{z^2} \mp, d_{x^2-y^2} \mp \}$, where we denoted up (down) spins with $+$ ($-$). 

\begin{equation}
    H_{d}^{\text{loc}} = \begin{pmatrix}
    \begin{array}{ccc|cc}
        \delta & \mp \frac{\lambda}{2} i & \frac{\lambda}{2}i & \pm\frac{\sqrt{3}\lambda}{2} & \mp\frac{\lambda}{2} \\[2pt]
        \pm\frac{\lambda}{2}i & \delta & \mp \frac{\lambda}{2} & -\frac{\sqrt{3}\lambda}{2}i & -\frac{\lambda}{2} i \\[2pt]
        -\frac{\lambda}{2}i & \mp\frac{\lambda}{2} & 0 & 0 & \mp\lambda i \\[2pt]
        \hline
        \rule{0pt}{1.25\normalbaselineskip}\pm\frac{\sqrt{3}\lambda}{2} & \frac{\sqrt{3}\lambda}{2}i & 0 & \Delta + \delta'  & 0 \\[2pt]
        \mp\frac{\lambda}{2} & \frac{\lambda}{2}i & \pm\lambda i & 0 & \Delta
    \end{array}
    \end{pmatrix}
    \label{eq:somatrix}
\end{equation}

One can notice from \autoref{eq:hamSO} 
that all the spin-orbit coupling elements within the $e_g$ sub-matrix are zero : no first-order spin-orbit interaction affects the $e_g$ states. 
Within the $t_{2g}$ states, at first order, it appears as if the orbital angular momentum was partially quenched from $l = 2$ to $l = 1$. When $\Delta \gg \lambda $ (separated by at least an order of magnitude), the $e_g$ and $t_{2g}$ submatrices can be assumed decoupled \cite{martinsj.phys.:condens.matter2017}. 
The $t_{2g}$ block will give rise to the \jeff = 3/2 and \jeff = 1/2 states, while the $e_g$ states will remain unaffected by SOC. Historically, this decoupling is refered to as the \textit{TP-equivalence approximation} \cite{sugano1970}.

The extracted value of the octahedral $\Delta$ and tetrahedral $\delta$ crystal-field splittings  in \bio\ and \bro\ are reported in \autoref{tab:param} too. By calculating the ratio $\lambda/\Delta$, one can evaluate the applicability of the TP-equivalence approximation. In \bio, this ratio is 10.1, placing it at (or close to) the limit of applicability of this approximation. In the case of \bro, due to a lower spin-orbit coupling constant, this ratio of 28.9 places it well within the range of applicability of this approximation.  

In the literature for layered iridates and rhodates, the TP-equivalence approximation has always been used to justify neglecting the $e_g$ orbitals in the effective low-energy models. However, we would like to point out that the hopping amplitude between the $d_{x^2-y^2}$ orbitals of neighboring Ir/Rh sites can significantly alter this picture by making the corresponding band to cross the Fermi level
: as a result, we rather propose a new interpretation of the TP-equivalence approximation for these compounds which would consist, instead of ignoring the $e_g$, in treating their coupling with the $t_{2g}$ manifold at the mean-field level. To support this statement, we will consider in the remainder of the article the full five-orbital local hamiltonian and its corresponding eigenstates : we will denote \jstar~the states mainly built out of the $t_{2g}$ manifolds and mark the spin-orbit $e_g$-like states with a tilde to differentiate them from the 'pure' $e_g$ states (for a more detailed derivation see \autoref{app:soc}). 

\subsection{Necessity of the full-\texorpdfstring{$d$}{d} manifold model}

The case of \bio\ and \bro\ strongly highlights the necessity of a five-orbital model description as soon as the DFT calculation. We claim that the previous five-orbital local hamiltonian \autoref{eq:somatrix} is actually the minimal model for all layered iridates and rhodates too. Our assessment is based on a set of arguments that are both theoretical and experimental :

\paragraph{Theoretically-based} For all layered iridates and rhodates, the full $d$ manifold is then the most natural and symmetric basis to express the hamiltonian : 
the "true" local eigenstates of \autoref{eq:somatrix} are rotationally-invariant contrary to the usual \jeff\ states built out of the three $t_{2g}$ orbitals only. Moreover, when calculating the Hubbard-Kanamori interaction, the full $d$-manifold is also the most natural basis, considering a restriction to the $t_{2g}$ manifold neglects parts of the orbital degrees of freedom that could influence the physics of the systems. In addition, the direct application of the TP-equivalence approximation has been recently questioned 
\cite{stamokostasphys.rev.b2018}, where the authors showed that in the case of 4$d$ and 5$d$ transition metals, even at large crystal-field splitting, the hybridization between the $e_g$ and the $t_{2g}$ states shouldn't be neglected.
\paragraph{Numerically-based} : The standard DFT calculations for \bio\ and \bro\ (see \autoref{fig:fig1}) highlight a crude proximity of the $e_g$ bands that crosses the Fermi level. This imposes the use of a  disentanglement scheme and even prevents a  building of Wannier functions for the $t_{2g}$ manifold. For \sio\ and Sr$_2$RhO$_4$, the distortions open a gap between the $e_g$ and $t_{2g}$ bands, but the building of Wannier functions for the $t_{2g}$ manifold alone may remain complicated. Moreover, standard tight-binding models for \sio\ and Sr$_2$RhO$_4$ have already highlighted the deviation from the \jeff\ picture due to the hybridization with the $e_g$ states : indeed, an additional term is usally introduced ad-hoc in the tight-binding of the $t_{2g}$ bands to get a better description of the upper bands, close to the $\Gamma$ point \cite{jin.phys.rev.B.2009,moutenet.phys.rev.B.2018,martinsphys.rev.mater.2018}. 
\paragraph{Experimentally-based} : 
Different X-Ray based experimental measurements on layered iridates (Resonant Inelastic X-Ray scattering\cite{ishiiphys.rev.b2011}, X-Ray Absorption Spectroscopy\cite{sohnscirep2016,moonphys.rev.b2006,morettisalaphys.rev.b2014}, Hard X-Ray Photoemission Spectroscopy\cite{yamasakiphys.rev.b2014}) reveal high-energy $d-d$ excitation and spectral features associated with transitions into the unnocupied $e_g$ states. Their combined sensitivity (to momentum resolved excitations in RIXS, orbital selective unoccupied states in XAS, and bulk electronic structure in HAXPES) expose a multi-orbital complexity that cannot be captured by the truncated $t_{2g}$ models. These observations motivate the need for theoretical frameworks that include the full $d$ manifold to consistently interpret experimental observations.

\section{The hybrid-DMFT (hMDFT) method}

\begin{figure}[htp]
    \centering 
    \includegraphics[width=0.95\linewidth]{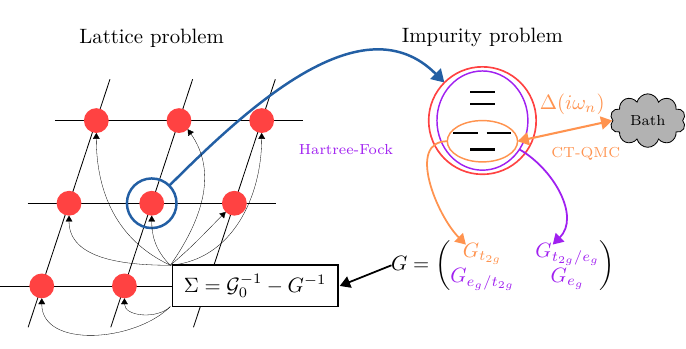}
    \caption{Schematic representation of the hybrid-DMFT approach for a $t_{2g}$/$e_g$ splitting of the local impurity model, as used in this article. The impurity problem is solved using CT-QMC part (in orange) for the $t_{2g}$ subspace while the rest is computed at a mean-field level (in purple).}
    \label{fig:ldmft}
\end{figure}

In a recent study of \bio~\cite{cassolphys.rev.b2024}, we showed that the effect of the correlations on the $e_g$ part of the spectrum was similar to a Hartree shift of their energies. 
This observation suggests that a simple mean-field treatment of this manifold could be sufficient to describe the low-energy physics of these systems. 
We argue that this results of the expected emptiness of the $e_g$ states and of the application of TP-equivalence approximation, which, in this system, would consist, instead of ignoring the $e_g$, in treating their coupling with the $t_{2g}$ manifold at the mean-field level.
As a consequence, we
developed a modification of the DMFT loop in which the $t_{2g}$ manifold is solved "accurately" and embedded in a mean-field treatment of the $e_g$ subspace and of the $t_{2g}-e_g$ coupling : we dubbed this approach \textit{hybrid-DMFT}. \autoref{fig:ldmft} shows an illustration of the hDMFT procedure, where the resulting Green function can directly be used in Dyson equation to close the DMFT loop. 

To be more precise, the usual DMFT loop \cite{georgesrev.mod.phys.1996} can be described as follows: 

\begin{enumerate}
    \item Pick a choice for a starting self-energy $\Sigma^0(i\omega_n)$
    \item Compute the local Green function : 
    \begin{equation}
        G_{\mathrm{loc}}^{N+1} = \sum_{\vk} \left[ i\omega_n + \mu - \hat{\mathrm{H}}_\vk - \Sigma^N(i\omega_n) \right]^{-1}
    \end{equation}
    \item Compute the Weiss field through Dyson equation : 
    \begin{equation}
        \left( \wf^{N+1}(i\omega_n) \right)^{-1} = \Sigma(i\omega_n)^N + \left( \gloc^{N+1}(i\omega_n) \right)^{-1}
    \end{equation}
    \item Use an impurity solver to compute the impurity Green function $\gimp^{N+1}$
    \item Define a new self-energy using Dyson equation: 
    \begin{equation}
        \Sigma_{\mathrm{imp}}^{N+1}(i\omega_n) = \left( \wf^{N+1}(i\omega_n) \right)^{-1} - \left( \gimp^{N+1}(i\omega_n) \right)^{-1}
    \end{equation}
    \item If the calculation is converged (i.e. if $\Sigma^{N+1} \approx \Sigma^{N}$), stop, else go back to step (2).
\end{enumerate}

In the fourth step, 
all the orbitals of the problem are treated on an equal footing. In the hDMFT approach, we will divide the manifold into layers that will be treated at different levels of theory. In the case of interest here,
we divide the local problem into two layers : 
The first layer will contain the complete $d$ manifold while the second will only contain the $t_{2g}$ subspace. The Weiss field can then be rewritten as : 
\begin{equation}
    \wf = \begin{pmatrix}
    \begin{array}{c|c}
    \wf^{t_{2g}} & \wf^{t_{2g}/e_g} \\ 
    \hline     
    \wf^{e_g/t_{2g}} & \wf^{e_g}
    \end{array}
    \end{pmatrix}
\end{equation}

Let's now call $G^{A}_{t_{2g}}$ the Green function obtained with an accurate impurity solver using the Weiss field $\wf^{t_{2g}}$. Let $G^{HF}$ be the Hartree-Fock Green function obtained using the density of the impurity Green function $G^{HF} = \left[ \wf^{-1} - \Sigma^{HF} \right]^{-1}$. The Hartree-Fock self-energy can be computed for a generic interaction term $\sum_{ijkl} U_{ijkl}\cre_i\cre_j\anh_k\anh_l$ as : 
\begin{equation}
    \Sigma^{\mathrm{HF}}_{ab} = 4 \sum_{ij} U_{aijb} \rho_{ij}
\end{equation}
where $\rho_{ij}$ is the local orbital occupancy matrix. 

The hDMFT Green function can then be written as : 

\begin{equation}
    G^{\mathrm{hDMFT}} = G^{HF} + \begin{pmatrix}
    \begin{array}{c|c} G^A_{t_{2g}} & 0 \\  \hline 0 & 0 \end{array} \end{pmatrix} - \begin{pmatrix}\begin{array}{c|c} G^{HF}_{t_{2g}} & 0 \\  \hline 0 & 0 \end{array} \end{pmatrix}
\end{equation}
where the last term is used to avoid the double counting of the Hartree-Fock part of the interaction in the resulting Green function. 
With this formulation, it is clear that only the $t_{2g}$ part of the Green function is computed with accurate method A, while all the other elements come from the Hartree-Fock Green function. 
This approach resembles DMFT+Hartree\cite{hansmannnewj.phys.2014,dattanatcommun2023, raiphys.rev.x2024}. The crucial difference in hDMFT is that the impurity problem is formulated for all the orbitals of interest, and the separation between the different methods is made at the impurity solver level.

\section{Results and discussion}

In the remainder of this article, the accurate method will be the Continuous Time Quantum Monte Carlo in the Hybridization expansion formalism (CT-HYB) \cite{sethcomput.phys.commun.2016}.
In all following DMFT calculations, the double counting will be considered isotropic and thus absorbed within the chemical potential. All further technical details  can be found in \autoref{app:computational}.

\subsection{Spectral functions}

First, we present the spectral functions (both momentum resolved and momentum integrated) for \bio~and \bro. The data presented in this section is not a direct output of the (h)DMFT calculation but has been obtained after an analytic continuation of the self-energy using the Maximum Quantum Entropy Method \cite{simphys.rev.b2018}.

    \paragraph{\texorpdfstring{\bio}{Ba2IrO4}}
    
\begin{figure*}[ht]
        \centering 
        \includegraphics[width=0.95\linewidth]{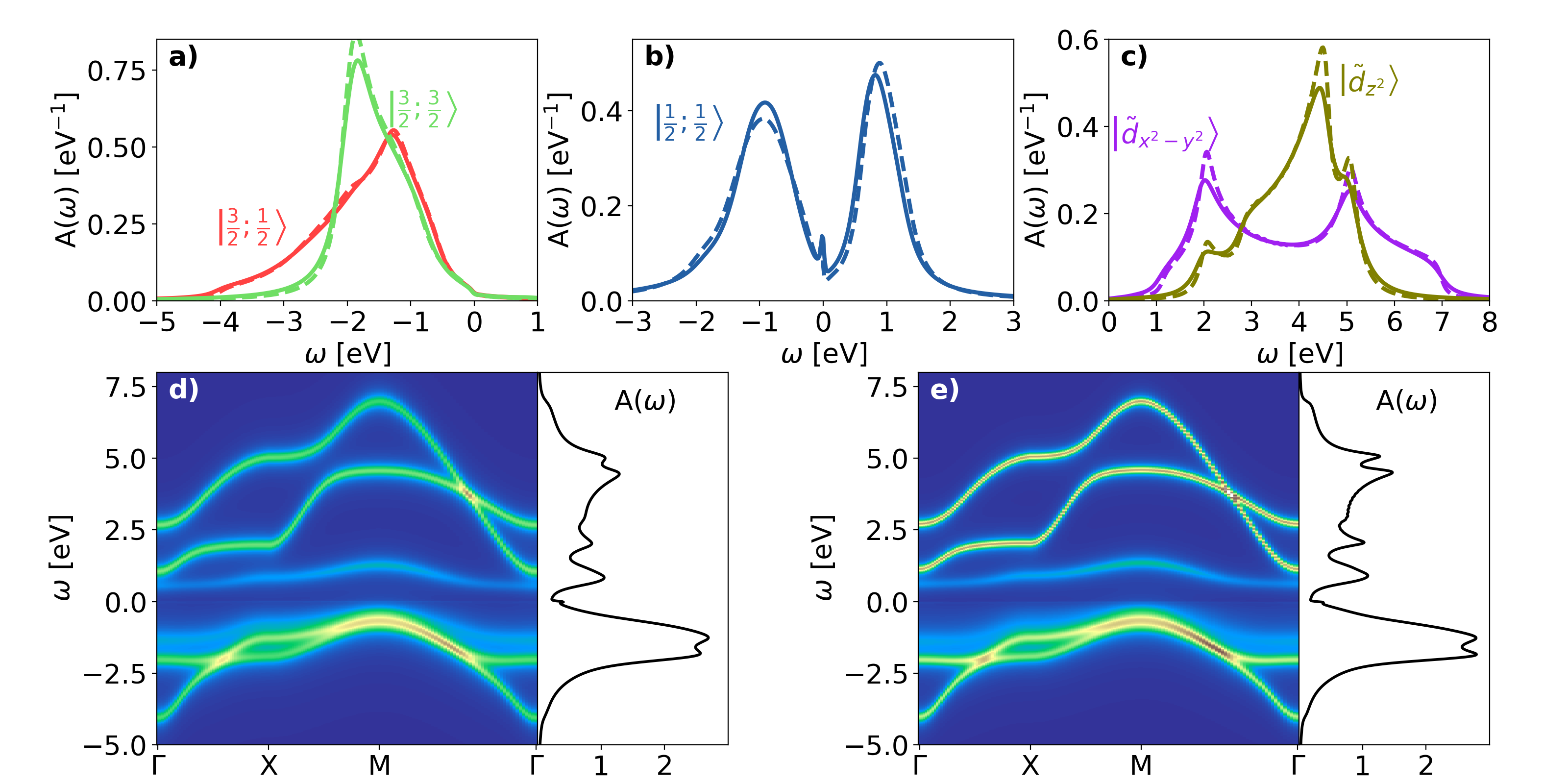}
        \caption{Orbital resolved (a-c) and momentum resolved (d-e) spectral functions computed for \bio~at $\beta=80$ eV$^{-1}$. a) DMFT (plain) and hDMFT (dashed) spectral functions of the filled part of the spectrum. b) DMFT (plain) and hDMFT (dashed) spectral functions of the \jstar=1/2 band. c) DMFT (plain) and hDMFT (dashed) spectral functions of the empty part of the spectrum. d) (e) show the DMFT (hDMFT) momentum resolved spectral function along the path $\Gamma$-X-M-$\Gamma$. The spectral functions were obtained using a modified Coulomb interaction $U = U_{\mathrm{cRPA}} + 0.3$ eV.}
        \label{fig:bio_spectral}
    \end{figure*}

    The spectral functions of \bio~obtained after analytic continuation of the self-energies are presented in \autoref{fig:bio_spectral}. 

    Following 
    our previous work on the three-band model \cite{cassolphys.rev.b2024}, we increase the value of the Coulomb matrix by 0.3~eV from the computed cRPA value in order to obtain an insulating solution. 
    With the set of parameters summarized in \autoref{tab:param}, the DMFT (and hDMFT) solution is however still metallic, albeit with a small weight at the Fermi level ($<$ 0.15 eV$^{-1}$) compared to the Hubbard bands ($\sim$ 0.4 eV$^{-1}$) as shown in pannel b).
    This feature is remnant of a quasiparticle peak, which is typical for a DMFT solution of the Hubbard model close to the metal-to-insulator transition. 

    In the (h)DMFT solution, the two \jstar=3/2 bands are completely filled (see panel a)), leaving the \jstar=1/2 band half-filled. This is exactly the \jeff=1/2 picture that is used to describe this system \cite{mosernewj.phys.2014}  and this is also consistent with the standard picture of \sio\ with a spin-orbital polarization enhanced by local correlation \cite{martinsphys.rev.lett.2011}
    
    As already mentionned in Ref. \cite{cassolphys.rev.b2024} 
    local correlations induce a spin-polarization of the system enhancing the differentiation between \jstar~and $\tilde{e}_g$ manifolds and
    the effect of the correlation on the $\tilde{e}_g$ bands appears to be primarly a Hartree shift. This assumption is now corroborated with the use of the hDMFT treatment that places these bands at the same position as the DMFT calculation shown in panel c). 
    
    Within the \jstar~subspace, the root-mean-square error (RMSE) computed between the DMFT and hDMFT spectral functions along the $\vK{}$ path is always below 7 meV$^{-1}$ while it can go to 15 meV$^{-1}$ within the $\tilde{e}_g$ subspace. 
    As hDMFT follows the assumption that the dynamical fluctuations can be neglected within the $\tilde{e}_g$ subspace, a higher error was expected. 
    Its small value still emphasizes that this assumption was adequate for this material. As we can observe in \autoref{fig:bio_spectral} panel c) to e), the main differences between the DMFT and hDMFT spectral functions resides in the sharpness of the features of the $\tilde{e}_g$ bands and their evenness along the $\vK{}$ path. 
    This is entirely due to the mean-field treatment of these bands, for which the only broadening appears during the analytic continuation, whereas the broadening of these bands within the DMFT scheme also includes an (albeit small) imaginary part of the self-energy. 

    \paragraph{\texorpdfstring{\bro}{Ba2RhO4}}

    In \autoref{fig:bro_spectral}, we report the spectral functions of \bro~obtained after analytic continuation of the self-energies. 
    Contrary to \bio, \bro~is clearly in a metallic state with a well-defined quasiparticle peak, which is consistent with available experimental measurements \cite{kurataphys.rev.materials2021}. 
    The DFT orbital occupancy $n_{j_{\mathrm{eff}}^*=1/2} = 1.53$, $n_{j^*_{\mathrm{eff}}=3/2;1/2} = 1.67$ and $n_{j^*_{\mathrm{eff}}=3/2;3/2} = 1.69$ has been redistributed by the electronic correlations, leading to a picture with a completely filled orbital and two partially occupied ones: 
    $n_{j^*_{\mathrm{eff}}=1/2} = 1.22$, $n_{j^*_{\mathrm{eff}}=3/2;1/2} = 1.98$ and $n_{j^*_{\mathrm{eff}}=3/2;3/2} = 1.78$. 
    and is very consistent with the filling
    of its distorted counterpart Sr$_2$RhO$_4$ \cite{martinsj.phys.:condens.matter2017}
    The $\tilde{e}_g$ bands are, as for \bio, completely empty.

    \begin{figure}[ht] 
        \centering 
        \includegraphics[width=0.95\linewidth]{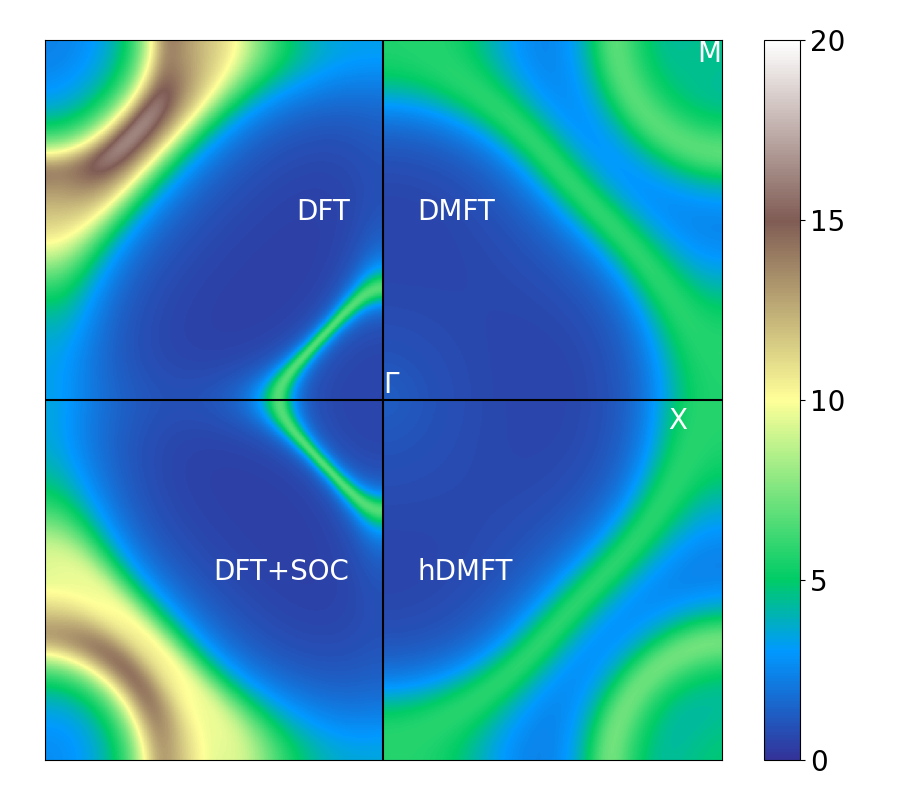} 
        \caption{Fermi surface of \bro~computed at different levels of theory. The central pocket in the DFT and DFT+SOC panels is attributed to the $\tilde{d}_{x^2-y^2}$ band that is pushed above the Fermi level within DMFT or hDMFT.
        }
        \label{fig:fs}
    \end{figure}

    The Fermi surface shown in \autoref{fig:fs} presents a remarkable agreement between DMFT and hDMFT. 
    There is however a noticeable difference with the Fermi surface obtained within DFT \cite{kurataphys.rev.materials2021}, where a noticeable pocket appears around the $\Gamma$ point, corresponding to the $\tilde{d}_{x^2-y^2}$ band. 
    We propose a new interpretation of the Fermi surface which is consistent with the interpretation of the Fermi surface of the sister compound Sr$_2$RhO$_4$. 
    Around the $M$ point, we observe an electron pocket labelled $\alpha$ of \jstar=3/2;3/2 character : that corresponds to the $\alpha$ pocket of Sr$_2$RhO$_4$ observed around the $\Gamma$ point due to the folding of the Brillouin zone. We measure its area to cover 7.7\% of the Brillouin zone.
    The second pocket labelled $\beta$ is a hole pocket of \jstar=1/2  character and is analogous to the $\beta_M$ pocket observed in Sr$_2$RhO$_4$, it covers 59.2\% of the Brillouin zone.
    
    \begin{figure*}[ht]
        \centering 
        \includegraphics[width=0.95\linewidth]{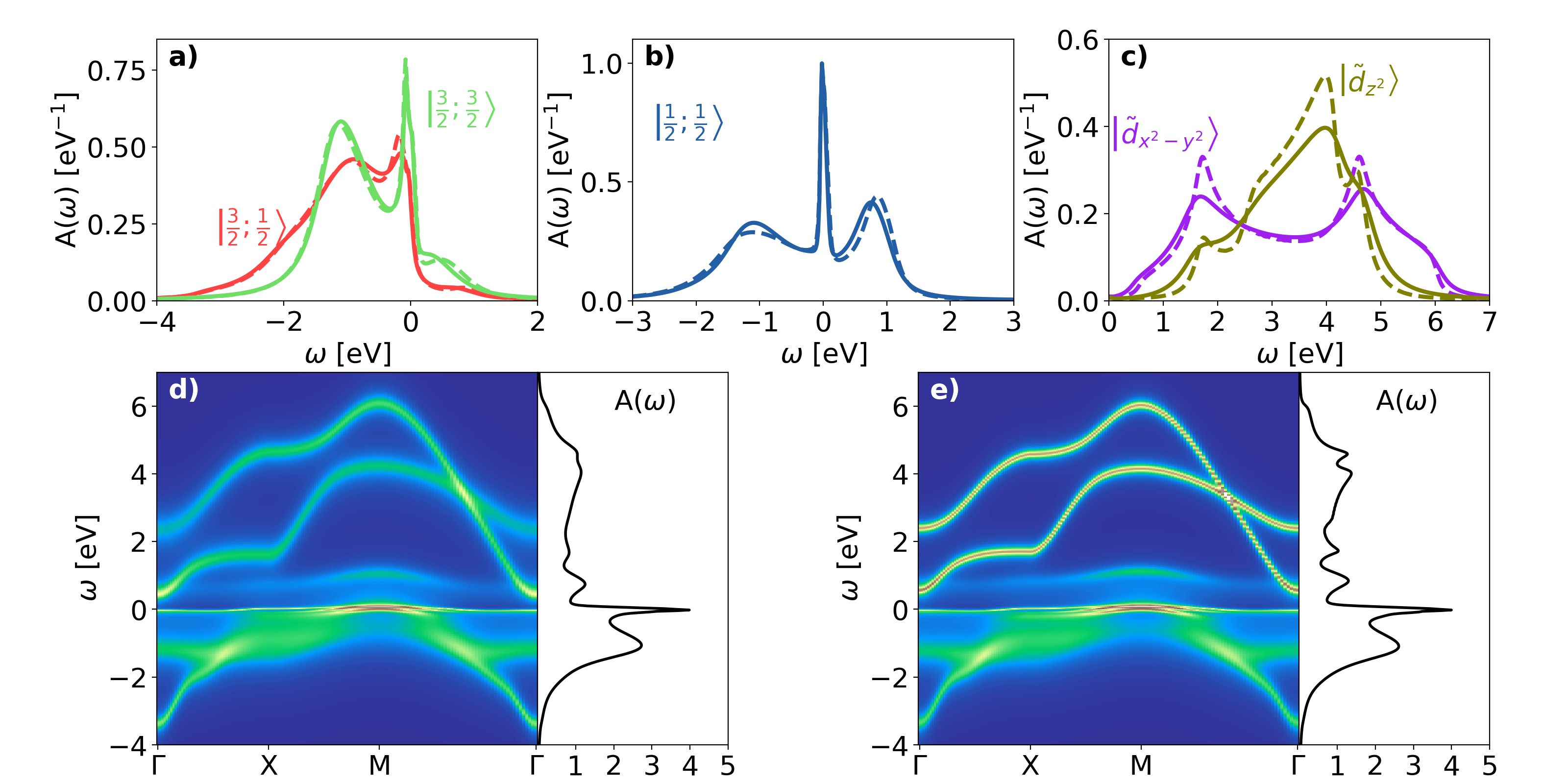}
        \caption{Orbital resolved (a-c) and momentum resolved (d-e) spectral functions computed for \bro~at $\beta=40$ eV$^{-1}$. a) DMFT (plain) and hDMFT (dashed) spectral functions of the \jstar=3/2 bands. b) DMFT (plain) and hDMFT (dashed) spectral functions of the \jstar=1/2 band. c) DMFT (plain) and hDMFT (dashed) spectral functions of the empty part of the spectrum. d) (e) show the DMFT (hDMFT) momentum resolved spectral function along the path $\Gamma$-X-M-$\Gamma$.}
        \label{fig:bro_spectral}
    \end{figure*}
    
    The RMSE between the DMFT and hDMFT spectral functions along the $\vK{}$ path is always below 4 meV$^{-1}$ within the \jstar~subspace while it can reach 19 meV$^{-1}$ within the $\tilde{e}_g$ subspace. As for \bio, a larger error within the $\tilde{e}_g$ subspace was expected. In \autoref{fig:bro_spectral} panel c), d) and e), one can clearly see the broadening of the $\tilde{e}_g$ orbitals in the DMFT calculation that hDMFT fails (by construction) to reproduce. 

\subsection{Influence on the self-energy}

In this section, we compare the effect of the hDMFT approximation to the imaginary part of the direct CTQMC output \jstar~self-energies and to the static limit of the self-energies of \bio~and \bro.

\begin{figure}[ht]
    \centering 
    \includegraphics[width=0.75\linewidth]{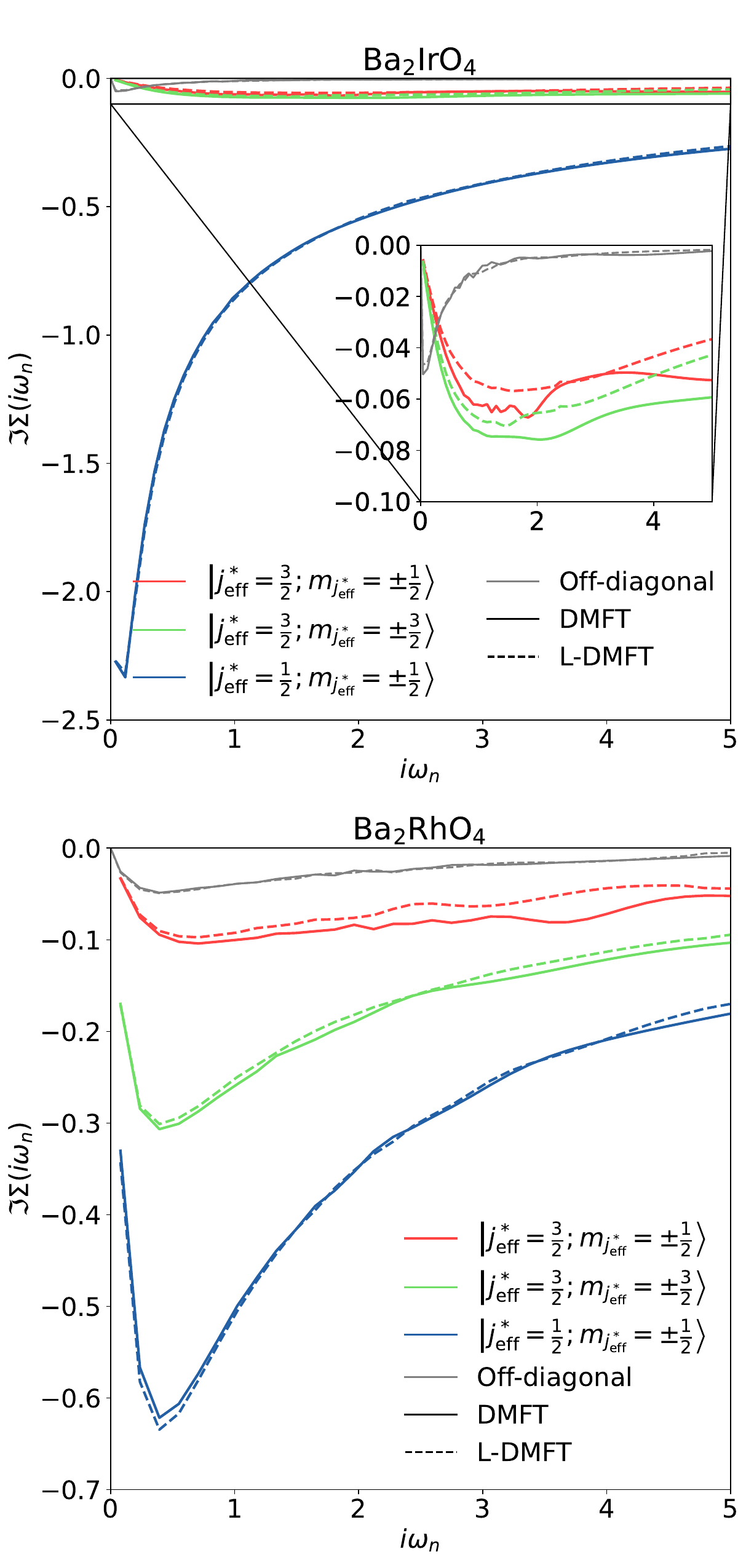}
    \caption{Comparison of the imaginary part of the DMFT (plain) and hDMFT (dashed) self-energies of \bio\ ($\beta=80$ eV$^{-1}$) and \bro ($\beta=40$ eV$^{-1}$) .}
    \label{fig:self}
\end{figure}

In \autoref{fig:self}, we report the converged self-energies in Matsubara space for \bio~and \bro~ for the DMFT (plain lines) and hDMFT (dashed lines) calculations. 

At low frequencies, the diagonal entries of the self energy present a remarkable agreement between the two methods. 
In gray, the only non-zero coupling within the \jstar~block, between the \jstar=1/2 and the \jstar=3/2;1/2 is shown. Despite being small, this coupling is also well reproduced within hDMFT. Across all the possible orbital indices and computed Matsubara frequencies, the RMSE was computed. For both \bio\ and \bro, the maximum value was 0.002~eV for the imaginary part of the self-energy and 0.008~eV for the real part of the self-energy.

\begin{table}[htp]
    \centering 
    \begin{tabular}{|c|c|c|c|c|c|}
    \hline 
    \parbox[c][20pt][c]{1.3cm}{\textbf{Ba$_2$IrO$_4$}}& $\left|\frac{3}{2};\pm \frac{1}{2} \right\rangle$ & $\left| \frac{3}{2};\pm \frac{3}{2} \right\rangle$ & $\left|\frac{1}{2};\pm \frac{1}{2} \right\rangle$ & $\left| \tilde{d}_{x^2-y^2} \right\rangle$ & $\left| \tilde{d}_{z^2} \right\rangle $ \\ 
    \hline 
    \hline 
    DMFT &  8.38 & 8.33 & 9.24 & 10.48 & 10.65 \\
    hDMFT & 8.37 & 8.32 & 9.24 & 10.48 & 10.65 \\
    \hline 
    \hline 
    \parbox[c][20pt][c]{1.5cm}{\textbf{Ba$_2$RhO$_4$}}& $\left|\frac{3}{2};\pm \frac{1}{2} \right\rangle$ & $\left| \frac{3}{2};\pm \frac{3}{2} \right\rangle$ & $\left|\frac{1}{2};\pm \frac{1}{2} \right\rangle$ & $\left| \tilde{d}_{x^2-y^2} \right\rangle$ & $\left| \tilde{d}_{z^2} \right\rangle $ \\ 
    \hline 
    \hline 
    DMFT & 6.02 & 6.16 & 6.53 & 7.71 & 7.81\\ 
    hDMFT & 6.02 & 6.15 & 6.53 & 7.70 & 7.80 \\
    \hline 
    \end{tabular}

    \caption{Comparison of $\Re \Sigma(i\omega_n) \rightarrow \infty$ between DMFT and hDMFT for \bio and~\bro}
    \label{tab:sigmainf}
\end{table}

In order to assess the accuracy of the hDMFT method concerning the static part, we compare in \autoref{tab:sigmainf} the real part of the self-energy evaluated at $i\omega_n \to \infty$. Again, for both materials the agreement between the two methods is remarkable, even here within the manifold treated at the mean-field level. 

\begin{table}[htp] 
    \centering
    \begin{tabular}{|c|c|c|c|c|c|}
    \hline 
    \parbox[c][20pt][c]{1.5cm}{}& $\left|\frac{3}{2};\pm \frac{1}{2} \right\rangle$ & $\left| \frac{3}{2};\pm \frac{3}{2} \right\rangle$ & $\left|\frac{1}{2};\pm \frac{1}{2} \right\rangle$ & $\left| d_{x^2-y^2} \right\rangle$ & $\left| d_{z^2} \right\rangle $ \\ 
    \hline 
    \hline 
    DMFT & 0.79 & 0.58 & 0.40 & 0.95 & 0.95\\ 
    hDMFT & 0.80 & 0.59 & 0.40 & 1.00 & 1.00 \\
    Sr$_2$RhO$_4$ (DMFT)\cite{martinsj.phys.:condens.matter2017} & / & 0.68 & 0.54  & / & / \\ 
    Sr$_2$RhO$_4$ (DMFT)\cite{ahnj.phys.:condens.matter2015} & 0.65 & 0.52 & 0.25 & / & / \\
    \hline 
    \end{tabular}
    \caption{Quasiparticle weight $Z = \left[ 1 - \frac{\partial \Im \Sigma(i\omega_n)}{\partial i\omega_n} \big|_{i\omega_n \to 0} \right]^{-1}$ computed from the DMFT and hDMFT self energies of \bro.}
    \label{tab:Z}
\end{table}

In \autoref{tab:Z}, we report the quasiparticle weights $Z = \left[ 1 - \frac{\partial \Im \Sigma(i\omega_n)}{\partial i\omega_n} \big|_{i\omega_n \to 0} \right]^{-1}$ computed from the self energies of the DMFT and hDMFT solutions. Here the agreement between DMFT and hDMFT is again remarkable on all the orbitals, and one can notice a small renormalization of the $\tilde{e}_g$ bands at the DMFT level that is missed by construction in the hDMFT solution. Moreover, we notice the strong renormalization of  the two partially-filled $j^*_{\mathrm{eff}=1/2;1/2}$ and $j^*_{\mathrm{eff}=3/2;3/2}$ bands. 
The renormalization of the quasiparticle mass shows similarities with the computed renormalization for Sr$_2$RhO$_4$ \cite{martinsj.phys.:condens.matter2017,ahnj.phys.:condens.matter2015} in which the $j_{\mathrm{eff}}=1/2$ band is the most renormalized. 
\subsection{Technical comparison}

Finally, we would like to point out one of the advantages of the hDMFT method over the DMFT method for a full five band calculation. 
In \autoref{table:speedup}, we present the speedup provided by the hDMFT calculation over the DMFT calculation for the two materials. For both \bio and \bro, the hDMFT calculation converged around 40 times faster than the DMFT calculation. 
We also report the average CT-QMC sign at convergence for both methods, and we notice that we get similar average signs for \bro. Interestingly, for Ba$_2$IrO$_4$, we see that hDMFT brings an improvement on the average sign at convergence. This improvement can be explained by the large sign problem arising in the calculations due to the strong spin-orbit interaction in this material. When effectively reducing the complexity of the CT-QMC from five to three bands, part of this sign problem is removed.

\begin{table}[h]
    \begin{tabular}{|c|c|c|}
    \hline
        & Ba$_2$IrO$_4$ & Ba$_2$RhO$_4$\\
        \hline 
       Speedup &  43.8 & 41.2\\
        \hline 
        DMFT average sign & 0.37 & 0.58\\
        hDMFT average sign & 0.53 & 0.60\\
        \hline 
    \end{tabular}
    \caption{Speedup of the total calculation time and average sign at convergence of hDMFT compared to DMFT for \bio~and \bro.}
    \label{table:speedup}
\end{table}

\section{Conclusion}

We have introduced hybrid Dynamical Mean-Field Theory (hDMFT), a controlled scheme 
that enables routine full $d$-manifold calculations for layered iridates and rhodates while maintaining near-quantitative accuracy in the low-energy physics. 
The method is motivated by the expected emptiness of the $\tilde{e}_g$ states and the TP-equivalence approximation : when the crystal-field splitting $\Delta$ is greater than the spin-orbit coupling strength $\lambda$, the $\tilde{e}_g$ states as well as their hybridization with the \jstar~ states can be treated at the Hartree-Fock level. 

A key achievement of this work is establishing the full five-orbital DMFT calculations for layered iridates and rhodates with realistic interaction strengths and spin-orbit coupling. 
These calculation were nonetheless essential to validate hDMFT and to definitely show that the $\tilde{e}_g$ states, are affected by a correlation-driven energy shift that places them well above the Fermi energy. 

Through the systematic benchmarking on \bio~and \bro, we have demonstrated that hDMFT reproduces the full five-orbital DMFT results with remarkable fidelity. For both materials, the \jstar~self-energies (including off-diagonal elements) did not show differences greater than the expected Monte Carlo noise. 

The physical features emerging from our calculations were able to clarify the role of the $\tilde{e}_g$ states in these materials. 
For \bio, we confirm the half-filled $j^*_{\mathrm{eff}}=1/2$ scenario with well separated $\tilde{e}_g$ states pushed $\sim 1.5$ eV above the Fermi level. 
For \bro, we find a filling consistent with its rhodate analog Sr$_2$RhO$_4$. Crucially, the $\tilde{d}_{x^2-y^2}$ pocket present in the DFT Fermi surface disappears upon inclusion of electronic correlations, substantially revisiting the interpretation of the electronic structure of this material. 

The computational advantage of hDMFT is shown substantial : we achieve a $\sim$40x speedup relative to the full five-orbital DMFT calculation for both materials.
Beyond wall-time savings, hDMFT improves the Monte Carlo sign problem for \bio~(average sign goes from 0.37 to 0.53), as reducing from five to three effective Monte Carlo orbitals removes part of the sign cancellations associated with strong spin-orbit coupling in the full $d$-manifold. 

The successful application to \bio\ and \bro\ suggests that hDMFT can also successfully be used for their more famous distorted counterparts. This will be the subject of future study. More generally, since
both Ir ($5d^5$) and Rh ($4d^5$) systems have been successfully described within hDMFT, at different correlation strengths and spin-orbit couplings, a wider range of applicability can be expected from other Ir or Rh-based compounds to other spin-orbit materials.

Finally, the hDMFT framework could be naturally extended to other multi-orbital problems where a subset of orbitals can be identified as weakly correlated (e.g. oxygen $p$ bands in charge-transfer materials).We however argue that 
a proper handling of the double-counting correction \cite{karolakjournalofelectronspectroscopyandrelatedphenomena2010,zhangphys.rev.lett.2016} will then be necessary . 

Several directions warrant future investigations. First, the treatment of double-counting correction deserves a systematic study.
Indeed, we used an isotropic double counting correction 
but orbital-dependent schemes are necessary for quantitative predictions of crystal-field enhancements and spectroscopic features,as it was shown in the case of the Fermi surface of \sro\ \cite{zhangphys.rev.lett.2016}. By extension, such improved double-counting may also be necessary for calculations of observables, like transport, where the role of $e_g$ states has recently been pointed out \cite{moserphys.rev.b2025}. Second, a systematic exploration of the ($\Delta$, $\lambda$, U) parameter space would establish rigorous applicability criteria for hDMFT beyond the materials studied in this article.

In summary, we have established hDMFT as a straightforward to implement and accurate method for incorporating the full $d$ manifold in strongly correlated spin-orbit materials. By recognizing that the TP-equivalence approximation justifies mean-field treatment rather than a complete neglect of the $e_g$ states, we resolve a longstanding tension between computational feasibility and physical completeness. The dramatic speedup achieved here opens the door to systematic 
studies that were previously computationally intractable, while maintaining the accuracy necessary for meaningful comparison with experiment. As a result, restriction to three-band low-energy models is no longer necessary and five-band DMFT calculation can then become the new standard to properly study these spin-orbit materials. 

\section*{Awknowledgments}

The authors acknowledge fruitful discussions with B. Lenz, S. Biermann, M. Aichhorn, J. Tomczak, P. Romaniello and E. Fromager which helped improve this work. Part of the calculations for this article were made possible by acces to the HPC resources of the CALMIP supercomputing center under the allocation No. P21048.

\appendix

\section{Computational details}
\label{app:computational}

Density functional theory (DFT) calculations were performed with the QuantumEspresso software package \cite{giannozzij.phys.:condens.matter2017,giannozzijphyscondensmatter2009}. We considered a plane wave basis set and optimized norm conserving Vanderbilt pseudopotentials \cite{hamannphys.rev.b2013} both in the scalar and full relativistic flavor for calculations with or without spin-orbit coupling. The DFT calculation were performed using the PBE exchange-correlation functional \cite{perdewphys.rev.lett.1996}.

For \bio, the crystal structure from Ref.~\cite{okabephys.rev.b2011} was taken as a starting point and the atomic positions were relaxed to a force $< 10^{-3}$ Ry/Bohr. The wavefunction energy cutoff was set to 90 Ry and a Monkhorst-Pack \cite{monkhorstphys.rev.b1976} regular $\vK{}$-points grid of 8x8x8 centered at $\Gamma$ was used for self-consistent calculations. 

For \bro, the crystal structure from Ref.~\cite{kurataphys.rev.materials2021} was taken as a starting point and the atomic positions were relaxed to a force $< 10^{-3}$ Ry/Bohr. The wavefunction energy cutoff was set to 92 Ry and a Monkhorst-Pack \cite{monkhorstphys.rev.b1976} regular $\vK{}$-points grid of 10x10x10 centered at $\Gamma$ was used for self-consistent calculations. 

For both materials, the tight-binding model was obtained via wannierization of the DFT band structure using maximally localized Wannier functions \cite{marzariphys.rev.b1997, souzaphys.rev.b2001} with the RESPACK software \cite{nakamuracomput.phys.commun.2021}. The outer energy window was set to [8.5 ev; 17.7 eV] ([8.2 eV; 16.1 eV]) for \bio~(\bro) and the frozen energy window was set to [10.35 eV; 14.2 eV] ([10.2eV; 14.0 eV]).

The Coulomb and exchange parameters were computed using the constrained Random Phase Approximation \cite{aryasetiawanphys.rev.b2004} method implemented in the RESPACK software \cite{nakamuracomput.phys.commun.2021}. A broadening of $0.1$ eV was considered for the Green function calculation and a maximum excitation energy of 200 eV was considered in the calculation of the polarization function for both materials. 

The Dynamical Mean Field Theory calculations were performed using the TRIQS library \cite{parcolletcomput.phys.commun.2015}, taking advantage of the DFTTools module \cite{aichhorncomput.phys.commun.2016} for the $\vK{}$ integrations. The continuous time quantum Monte Carlo impurity solver \cite{wernerphys.rev.lett.2006} formulated in the hybridization expansion formalism from the TRIQS library \cite{sethcomput.phys.commun.2016} was used as the DMFT solver and the numerically accurate hDMFT solver. 
For \bio, the Green function was first converged at $\beta=30$ using 28.8$\times 10^6$ (144 $\times 10^6$) Monte Carlo measurements for each iteration. The temperature was then lowered to $\beta=80$ where the Green function was converged using 144$\times 10^6$ (144 $\times 10^6$) Monte Carlo measurements for each iteration for the DMFT (hDMFT) calculation.
For \bro, the Green function was converged at $\beta=40$ using 72$\times 10^6$ Monte Carlo measurements for each iteration for the DMFT and hDMFT calculation. 

The analytic continuation was performed using the maximum quantum entropy method implemented in the MQEM code \cite{simphys.rev.b2018} using a smearing factor of 0.03 (0.05) eV for \bio~(\bro). 

In order to compute the spectral functions, a broadening of 0.1 eV was applied for both materials. The momentum integrated spectral function was computed using a 80x80x80 regular $\vK{}$-point grid for both materials.

\section{Low-energy models}
\label{app:models}

In this appendix, we describe the low energy models used to describe \bio~and \bro. 
In order to compare the spatial extent of the Wannier functions between the two systems, we define the following quantity : 
\begin{equation}
    \sigma = \sqrt{\frac{\Omega}{d^2_ {\mathrm{M-M}}}}
\end{equation}
where $\Omega$ is the spread of the Wannier function and $d_{\mathrm{M-M}}$ is the in-plane distance between two metallic centers. With this definition, $\sigma$ corresponds to the average in-plane spatial extent of the Wannier functions as a proportion of the metal-metal distance. 

\begin{table}
    \centering 
    \begin{tabular}{|c|c|c|c|c|c|}
    \hline
        $\sigma$ & $d_{xy}$ & $d_{xz}$ & $d_{yz}$ & $d_{x^2-y^2}$\\ 
        \hline
        \bio & 0.51 & 0.51 & 0.51 & 0.46 \\
        \hline
        \bro & 0.58 & 0.54 & 0.54 & 0.45 \\
        \hline
    \end{tabular}

    \caption{In-plane average spatial extent $\sigma$ of the Wannier function of the models used for \bio~and \bro.}
    \label{tab:spread}
\end{table}

In \autoref{tab:spread}, we show the value of $\sigma$ for the Wannier functions of interest of the models for \bio~and \bro. 
In both cases, we observe the $d_{x^2-y^2}$ Wannier funtion to have a smaller spatial extent than the $t_{2g}$ Wannier functions. 
This is due to the orientation of the $d_{x^2-y^2}$ orbital, pointing towards the oxygen atoms, which constrains it in space. 
For the case of \bio, interestingly, the three $t_{2g}$ orbitals share a similar spatial extent which is not the case for \bro~where the $d_{xy}$ Wannier function has a larger spread. 
This suggests \bio~to be more isotropic than \bro. 

\begin{table}
    \centering 
    \begin{tabular}{|c|c|c|c|c|c|c|} 
        \hline
        & $t_{xy}$ & $t_{yz}$ & $t_ {xz}$ & $t_{x^2-y^2}$ & $t_{z^2}$ & $t_{x^2-y^2/z^2}$ \\
        \hline 
        \bio & -0.367 & -0.065 & -0.246 & -0.702 & -0.213 & 0.371\\ 
        \hline 
        \bro & -0.289 & -0.060 & -0.172 & -0.641 & -0.193 & 0.352\\ 
        \hline 
    \end{tabular}
    \caption{Nearest-neighbor hopping (in the $x$ direction) between the $d$ Wannier orbitals of \bio~and\bro.}
    \label{tab:hop}
\end{table}

\autoref{tab:hop} shows the nearest-neighbor hopping of the Wannier hamiltonian for \bio~and \bro.
In both materials, the hopping intensities follow the same trend. In \bro, the hoppings are systematically lower than in \bio, except for the $t_{yz}$ hopping, which appears to be similar in both systems. 

\begin{table}[ht]
        \begin{tabular}{c|ccccc}
            \hline
            \hline
            U (eV) & $d_{xy}$ & $d_{yz}$ & $d_{xz}$ & $d_{x^2-y^2}$ & $d_{z^2}$ \\
            \hline
            $d_{xy}$ & 2.03 & 1.43 & 1.43 & 1.85 & 1.47 \\
            $d_{yz}$ & 1.43 & 2.04 & 1.54 & 1.57 & 1.78 \\
            $d_{xz}$ & 1.43 & 1.54 & 2.04 & 1.57 & 1.78 \\
            $d_{x^2-y^2}$ & 1.85 & 1.57 & 1.57 & 2.56 & 1.63 \\
            $d_{z^2}$ & 1.47 & 1.78 & 1.78 & 1.63 & 2.44 \\
            \hline
            \hline
        \end{tabular}
        \begin{tabular}{c|ccccc}
            \hline
            \hline
            J (eV) & $d_{xy}$ & $d_{yz}$ & $d_{xz}$ & $d_{x^2-y^2}$ & $d_{z^2}$ \\
            \hline
            $d_{xy}$ & 0.00 & 0.21 & 0.21 & 0.21 & 0.27 \\
            $d_{yz}$ & 0.21 & 0.00 & 0.21 & 0.25 & 0.21 \\
            $d_{xz}$ & 0.21 & 0.21 & 0.00 & 0.25 & 0.21 \\
            $d_{x^2-y^2}$ & 0.21 & 0.25 & 0.25 & 0.00 & 0.37 \\
            $d_{z^2}$ & 0.27 & 0.21 & 0.21 & 0.37 & 0.00 \\
            \hline
            \hline
        \end{tabular}
        \caption{Coulomb and exchange parameters obtained for the five-band model of \bro.}
        \label{tab:bro_coulomb}
\end{table}

In \autoref{tab:bio_coulomb} and \autoref{tab:bro_coulomb}, we report the computed cRPA values for the $\omega=0$ local Coulomb interaction (U) and Hund coupling (J). For the Coulomb interaction, we observe an opposite effect as the one observed for the spread : the $t_{2g}$ part of the Coulomb matrix  shows a greater anisotropy for \bro~than for \bio.
For \bio, the value we find are consistent with the litterature \cite{aritaphys.rev.lett.2012, mosernewj.phys.2014}. The value of the local Coulomb matrix are greater than the average value of the in plane nearest-neighbor non-local Coulomb repulsion ($V = 0.71$ eV) which we neglect in the DMFT approximation.
For \bro, the diagonal $t_{2g}$ elements of the Coulomb matrix are similar, but the anisotropy of the system is still present in the off diagonal elements which differ. The values obtained are consistent with cRPA interaction values computed for Sr$_2$RhO$_4$ without distortions \cite{martinsj.phys.:condens.matter2017}.

\begin{table}[ht]
        \centering
        \begin{tabular}{c|ccccc}
        \hline
        \hline
        U (eV) & $d_{xy}$ & $d_{yz}$ & $d_{xz}$ & $d_{x^2-y^2}$ & $d_{z^2}$ \\
        \hline
        $d_{xy}$ & 2.41 & 1.73 & 1.73 & 2.16 & 1.73 \\
        $d_{yz}$ & 1.73 & 2.31 & 1.78 & 1.83 & 2.01 \\
        $d_{xz}$ & 1.73 & 1.78 & 2.31 & 1.83 & 2.01 \\
        $d_{x^2-y^2}$ & 2.16 & 1.83 & 1.83 & 2.81 & 1.86 \\
        $d_{z^2}$ & 1.73 & 2.01 & 2.01 & 1.86 & 2.64 \\
        \hline
        \hline
        \end{tabular}
        \begin{tabular}{c|ccccc}
        \hline
        \hline
        J (eV) & $d_{xy}$ & $d_{yz}$ & $d_{xz}$ & $d_{x^2-y^2}$ & $d_{z^2}$ \\
        \hline
        $d_{xy}$ & 0.00 & 0.22 & 0.22 & 0.22 & 0.27 \\
        $d_{yz}$ &  0.22 & 0.00 & 0.22 & 0.26 & 0.21\\
        $d_{xz}$ &  0.22 & 0.22 & 0.00 & 0.26 & 0.21 \\
        $d_{x^2-y^2}$ &  0.22 & 0.26 & 0.26 & 0.00 & 0.36\\
        $d_{z^2}$ & 0.27 & 0.21 & 0.21 & 0.36 & 0.00 \\
        \hline
        \hline
        \end{tabular}
        \caption{Coulomb and exchange parameters obtained for the five-band model of \bio.}
        \label{tab:bio_coulomb}
\end{table}

\section{\texorpdfstring{Definition of the \jeff~states}{Definition of the jeff states}}
\label{app:soc}

Let's rewrite the local Hamiltonian in the orbital $d$ basis, in presence of spin-orbit coupling ($\lambda$), octahedral ($\Delta$) and tetragonal ($\delta$) crystal field splitting in the basis $\{ d_{xz} \pm, d_{yz} \pm, d_{xy} \mp, d_{z^2} \mp, d_{x^2-y^2} \mp \}$: 

\begin{equation}
    H_{d}^{\text{loc}} = \begin{pmatrix}
    \begin{array}{ccc|cc}
        \delta & \mp \frac{\lambda}{2} i & \frac{\lambda}{2}i & \pm\frac{\sqrt{3}\lambda}{2} & \mp\frac{\lambda}{2} \\[2pt]
        \pm\frac{\lambda}{2}i & \delta & \mp \frac{\lambda}{2} & -\frac{\sqrt{3}\lambda}{2}i & -\frac{\lambda}{2} i \\[2pt]
        -\frac{\lambda}{2}i & \mp\frac{\lambda}{2} & 0 & 0 & \mp\lambda i \\[2pt]
        \hline
        \rule{0pt}{1.25\normalbaselineskip}\pm\frac{\sqrt{3}\lambda}{2} & \frac{\sqrt{3}\lambda}{2}i & 0 & \Delta + \delta'  & 0 \\[2pt]
        \mp\frac{\lambda}{2} & \frac{\lambda}{2}i & \pm\lambda i & 0 & \Delta
    \end{array}
    \end{pmatrix}
\end{equation}

The eigenstates of this Hamiltonian can be perturbatively expanded when the ratio $\lambda/\Delta$ is large :

\begin{align}
        \ket{\tilde{d}_{z^2}} &= \ket{d_{z^2}} + \mathcal{O}\left( \frac{\lambda}{\Delta}\right) \\
        \ket{\tilde{d}_{x^2-y^2}} &= \ket{d_{x^2-y^2}} + \mathcal{O}\left(\frac{\lambda}{\Delta}\right) \\
        \ket{\frac{1}{2}, \pm \frac{1}{2}} &=  \frac{\cos \theta}{\sqrt{2}} \left(  i\ket{d_{xz},\mp} \mp  \ket{d_{yz}, \mp} \right) + \sin  \theta \ket{d_{xy},  \pm} +  \mathcal{O}\left( \frac{\lambda}{\Delta} \right)\\
        \ket{\frac{3}{2}, \mp \frac{1}{2}} &= \frac{\sin \theta}{\sqrt{2}}  \left( i\ket{d_{xz}, \mp} \mp \ket{d_{yz},  \mp} \right) - \cos\theta \ket{d_{xy}, \pm}  + \mathcal{O}\left( \frac{\lambda}{\Delta} \right) \\
        \ket{\frac{3}{2}, \pm \frac{3}{2}} &= \frac{1}{\sqrt{2}} \left(i\ket{d_{xz}, \pm} \mp \ket{d_{yz}, \pm} \right) + \mathcal{O}\left( \frac{\lambda}{\Delta} \right)
    \end{align}

    Where $\pm$ denotes up or down spin, and $\theta$ is a mixing angle that  depends on $\lambda$  and  $\delta$  such that :

\begin{equation}
    \sin 2\theta  = \frac{\sqrt{2}  \lambda}{\sqrt{2\lambda^2 + (\delta + \frac{\lambda}{2})^2}}
        \label{eq:mix}
\end{equation}

The ket states represent the \jstar~states, where the star is used to differentiate them from the usual \jeff~states. Similarly, the tilde on the $\tilde{e}_g$ states is here to show that they are different from the pure $e_g$ states. With the parameters used to describe \bro~and \bio, the deviation from the "pure" \jeff ~and $e_g$ states is less than 1\%.
When the tetragonal splitting $\delta$ becomes 0, we recover the ideal \jeff~picture where the \jeff=1/2 bears equal weight on the three $t_{2g}$ orbitals.

\clearpage
\bibliography{references}

\end{document}